\documentclass{wileyarticle}

\articletype{}%

\usepackage{lipsum}

\begin{document}

\title{The Magnetic Field Distribution in Strongly Magnetized Neutron Stars}

\author[1]{V. Dexheimer}

\author[3]{B. Franzon}

\author[3]{R. O. Gomes}

\author[4]{R. L. S. Farias}

\author[5]{S. S. Avancini}

\author[3]{S. Schramm}

\authormark{V. Dexheimer \textsc{et al}}

\address[1]{\orgdiv{Department of Physics}, \orgname{Kent State University}, \orgaddress{\state{Kent OH}, \country{USA}}}

\address[3]{\orgdiv{FIAS}, \orgname{Johann Wolfgang Goethe University}, \orgaddress{\state{Frankfurt}, \country{DE}}}

\address[4]{\orgdiv{Departamento de F\'isica}, \orgname{Universidade Federal de Santa Maria}, \orgaddress{\state{Santa Maria, RS}, \country{Brazil}}}

\address[5]{\orgdiv{Departamento de F\'isica}, \orgname{Universidade Federal de Santa Catarina}, \orgaddress{\state{Florian\'opolis, SC}, \country{Brazil}}}

\corres{V. Dexheimer, \email{vdexheim@kent.edu}\\ 
}


\begin{abstract}%
In this work, we expand on a previously reported realistic calculation of the magnetic field profile for the equation of state inside strongly magnetized neutron stars. In addition to showing that magnetic fields increase quadratically with increasing baryon chemical potential of magnetized matter (instead of exponentially, as previously assumed), we show here that the magnetic field increase with baryon number density is more complex and harder to model. We do so by the analysis of several different realistic models for the microscopic description of matter in the star (including hadronic, hybrid and quark models) combined with general relativistic solutions by solving Einstein-Maxwell's field equations in a self-consistent way for stars endowed with a poloidal magnetic field.
\end{abstract}



\maketitle

\section{Introduction}

In order to study the effects of magnetic fields in the equation of state of neutron stars, a profile for
the strength of the field as a function of chemical potential or density must be provided. If one has access to a code that calculates general relativistic solutions in the presence of a magnetic field by solving Einstein-Maxwell's field equations in a self-consistent way, this can be easily achieved.  Alternatively, ad hoc profiles for the magnetic field have been provided and used by the nuclear physics community for the past two decades.

The first of these ad hoc profiles was suggested in Ref.~\cite{Bandyopadhyay:1997kh}

\begin{eqnarray}
B^*(n_B/n_0)=B_{\rm{surf}}+B_0\left[1-e^{-\beta(n_B/n_0)^\gamma}\right],
\label{1}
\end{eqnarray}
with typical choices of constants $\beta=0.01$ and $\gamma =3$. In this case, the magnetic field increases exponentially  from a value $B_{\rm{surf}}$ at zero density to a value $B_{\rm{surf}}+B_0$ at asymptotically high densities. 

This profile was subsequently used in approximately one hundred publications, among which the most cited ones are 
\cite{Menezes:2008qt,Menezes:2009uc,Barkovich:2004jp,Dexheimer:2011pz}. An improvement
over this formulation assumes a field profile as a function of baryon chemical potential $\mu_B$ (\cite{Dexheimer:2011pz}). Although the second ansatz does not suffer from spurious jumps in the strength of the magnetic field in the presence of first order phase transitions (such as the deconfinement to quark matter), it is still not correct. As already pointed out in Ref.~\cite{Menezes:2016wbw}, ad hoc formulas for magnetic field profiles in neutron stars such as Eq.~\ref{1} do not fulfill Maxwell's equations and, therefore, are incorrect.

In this work, we calculate the magnetic field distribution in the polar stellar direction and translate it to be a function of microscopic thermodynamical quantities, the baryon chemical potential and for the first time the baryon number density. In order to do so, the macroscopic structure of the star must be obtained from the solution of Einstein-Maxwell equations. Only in this way, can we ensure that the magnetic field profile in a star respects the Einstein-Maxwell field equations. In order to make our analysis as general as possible, we make use of three equations of state for the microscopic description of neutron stars with different matter composition: hadronic, hybrid and quark. They represent state-of-the-art approaches that fulfill current nuclear and astrophysical constraints, such as the prediction of massive stars.

\section{Formalism}

The first model was obtained from Refs.~\cite{Gomes:2014aka,Gomes:2014dka} and it will be referred to as ``G-model''. It is a hadronic model that simulates many-body forces among nucleons by non-linear self-couplings that come from a field dependence of the interactions. The second model was obtained from Refs.~\cite{Dexheimer:2009hi,Dexheimer:2011pz} and it will be referred to as ``D-model". It includes nucleons, hyperons and quarks in a self-consistent approach and reproduces chiral symmetry restoration and deconfinement at high densities. The third model was obtained from Ref.~\cite{Hatsuda:1994pi} and it will be referred to as ``H-model". It is a version of the three-flavor NJL model that includes a repulsive vector-isoscalar interaction, which is crucial for the description of astrophysical data.

For the general-relativistic formalism to describe the macroscopic features of magnetic neutron stars, we use the LORENE C++ class library for numerical relativity (\cite{Bonazzola:1993zz,Bocquet:1995je}), which determines equilibrium configurations by solving the Einstein-Maxwell's field equations in spherical polar coordinates assuming a poloidal magnetic field configuration. In this approach, the field is produced self-consistently by a macroscopic current, which is a function of the stellar radius, polar angle theta, and dipole magnetic moment $\mu$ for each equation of state. The dipole magnetic moments shown in this work were chosen to reproduce a distribution with a central stellar magnetic field close to the upper limit of the code $\sim10^{18}$ G and one to reproduce a surface magnetic field of $\sim10^{15}$ G, the maximum value observed on the surface of a star (\cite{Melatos:1999ji}).

\begin{figure}[t!]
\SPIFIG{\includegraphics[trim={0 0 1.cm 2.7cm},clip,width=270pt]{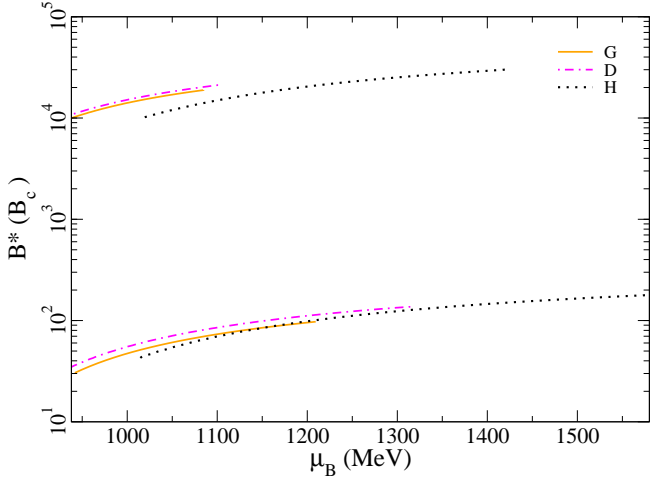}}{\caption{(Color online) Magnetic field profile in the polar direction in a $M_B=2.2$ M$_\odot$ star as a function of baryon chemical potential obtained for the three equation of state models R, D and H. Each of these profiles are shown for dipole magnetic moments $\mu=3\times10^{32}$ Am$^2$ (curves on the top) and $\mu=1\times10^{30}$ Am$^2$ (curves on the bottom). \label{newfig3}}}
\end{figure}

We calculate the equation of state within the microscopic models without magnetic field effects, as we have already shown in Refs.~\cite{Dexheimer:2016yqu,rosana} that they do not affect significantly the stellar magnetic field distribution. Then, in a second step, through the solution of Einstein's equations coupled with Maxwell's equations, we determine the magnetic field profile in an individual star, and then translate that to a field profile for the microscopic equation of state of each model. Later, we discuss a generalization to one profile by averaging the results from the different models.

\section{Results}

Figure~\ref{newfig3} shows the magnetic field distribution in the stellar polar direction for a $M_B=2.2$ M$_\odot$ star translated into the microscopic quantity baryon chemical potential. See Ref.~\cite{Dexheimer:2016yqu} for figures showing the magnetic field as a function of stellar radius. The top curves of the figure are magnetic field profiles in the stellar polar direction for a higher dipole magnetic moment, while the bottom curves are profiles for a lower value of the dipole magnetic moment. The main conclusion from this figure is that different equation of state models show different magnetic field strengths, but the respective profiles have approximately the same shape (when taking into account the logarithmic scale). The shape of the profiles obtained from the solution of Einstein-Maxwell's equations is well fit by a quadratic polynomial (and not exponential function, as suggested by ad hoc profiles). This allows us to fit one profile using the average of the different equation of state models. It depends only on the baryon chemical potential $\mu_B$ and on the value chosen for the dipole magnetic moment $\mu$
\begin{eqnarray}
B^*(\mu_B)=\frac{(a + b \mu_B + c \mu_B^2)}{B_c^2} \ \mu,
\label{3}
\end{eqnarray}
with $\mu_B$ given in MeV and $\mu$ in Am$^2$ in order to produce $B^*$ in units of the critical field for the electron $B_c=4.414\times 10^{13}$~G. The coefficients $a$, $b$, and $c$ given in Table~\ref{table}.

\begin{figure}[t!]
\SPIFIG{\includegraphics[trim={0 0 1.cm 2.7cm},clip,width=270pt]{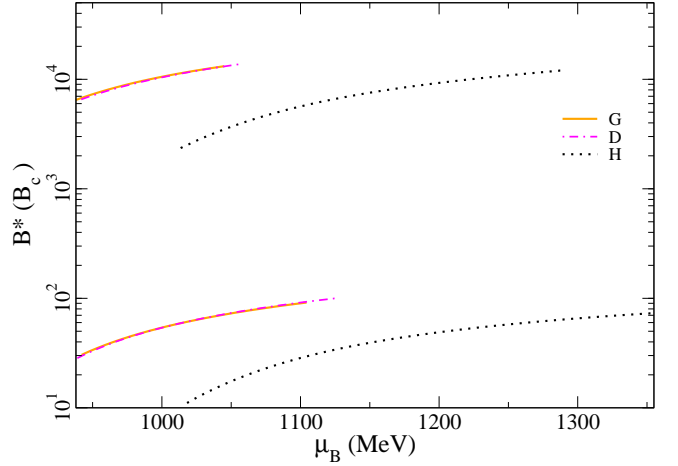}}{\caption{(Color online) Same as Fig.~\ref{newfig3} but for a $M_B=1.6$ M$_\odot$ star with dipole magnetic moments $\mu=2\times10^{32}$ Am$^2$ (curves on the top) and $\mu=1\times10^{30}$ Am$^2$ (curves on the bottom). Some of the curves overlap.\label{newfig4}}}
\end{figure}

In Fig.~\ref{newfig4}, we repeat the calculations for a $M_B=1.6$ M$_\odot$ star again for different dipole magnetic moments. Once more, each magnetic field profile has the same shape (when taking into account the logarithmic scale). In this case, the parameters of the profile fit in Eq.~(\ref{3}) are again given in Table~\ref{table}, where from the values of the parameter ``c" it can be seen that the profiles for a larger star give on average a slightly more linear fit. Note that for a less massive and less compact stars, all equations of state that contain baryons reproduce very similar results. This stems from the fact that they were fitted to reproduce nuclear physics constraints and the central densities in such stars do not reach values much larger than nuclear saturation density.

For a detailed comparison between the results from Figs.~\ref{newfig3} and \ref{newfig4} and ad hoc exponential profiles, see Ref.~\cite{Dexheimer:2016yqu}. Clearly, none of the ad hoc exponential profiles coincide with our results (except maybe for one point), even the ad hoc profiles that were chosen to match our field strengths on the surface of the star and at asymptotically high chemical potentials. 

\begin{table}[t!]%
\processtable{\label{table} Quadratic fit coefficients $a$, $b$, and $c$ for Eq.~{3} calculated for different baryonic mass stars.}
{\tabcolsep=0pt%
\begin{tabular*}{\columnhsizevalue}{@{\tabindent\extracolsep\fill}lc|ccc@{\extracolsep\fill\tabindent}}
\hline\\[-1.63em]
&\Bigg.$M_B$ (M$_\odot$) \ \ & $a$ $\left(\frac{\rm{G}^2}{\rm{A m}^2}\right)$ & $b$ $\left(\frac{\rm{G}^2}{\rm{A m}^2 \rm{MeV}}\right)$ & $c$ $\left(\frac{\rm{G}^2}{\rm{A m}^2 \rm{MeV}^2}\right)$\ \ \ \ \\
\hline
&$2.2$ & $-7.69\times10^{-1}$ & $1.20\times10^{-3}$ & $-3.46\times10^{-7}$\ \ \ \ \\
&$1.6$ & $-1.02$ & $1.58\times10^{-3}$ & $-4.85\times10^{-7}$\ \ \ \ \\
\botrule
\end{tabular*}}{\begin{tablenotes}
\end{tablenotes}}
\end{table}

\begin{figure}[t!]
\SPIFIG{\includegraphics[trim={0 0 1.cm 2.7cm},clip,width=270pt]{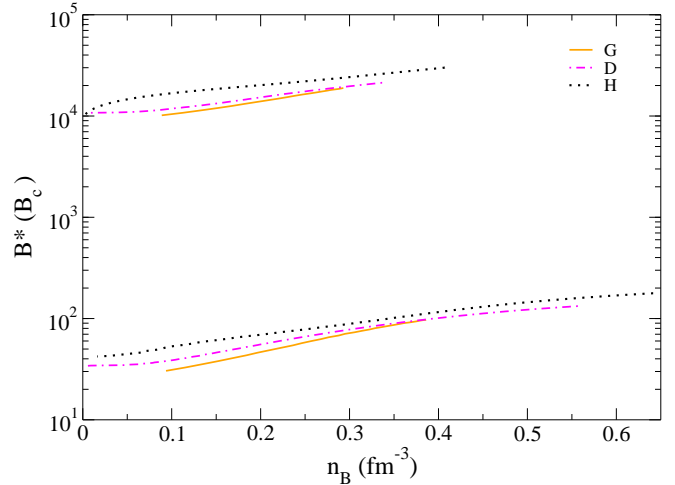}}{\caption{(Color online) Magnetic field profile in the polar direction in a $M_B=2.2$ M$_\odot$ star as a function of baryon number density obtained for the three equation of state models R, D and H. Each of these profiles are shown for dipole magnetic moments $\mu=3\times10^{32}$ Am$^2$ (curves on the top) and $\mu=1\times10^{30}$ Am$^2$ (curves on the bottom). \label{rhohigh}}}
\end{figure}

\begin{figure}[t!]
\SPIFIG{\includegraphics[trim={0 0 1.cm 2.7cm},clip,width=270pt]{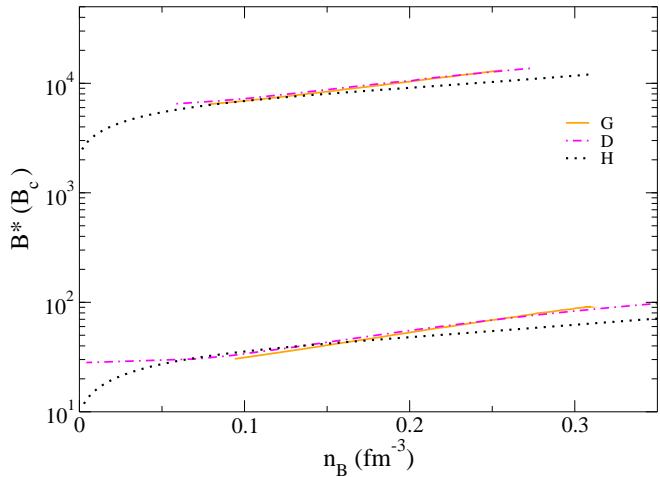}}{\caption{(Color online) Same as Fig.~\ref{rhohigh} but for a $M_B=1.6$ M$_\odot$ star with dipole magnetic moments $\mu=2\times10^{32}$ Am$^2$ (curves on the top) and $\mu=1\times10^{30}$ Am$^2$ (curves on the bottom).\label{rholow}}}
\end{figure}

Next, we focus on the discussion of magnetic field distributions as a function of baryon number density. This is shown is Figs.~ \ref{rhohigh} and \ref{rholow} for stars with different baryon masses (and in each figure for different dipole magnetic moments). It can immediately be seen that all the curves have different shapes. In Fig.~\ref{rhohigh}, it can be clearly seen that the curves for the ``G" EoS model looks quadratic, while the others are better fit by a quartic polynomial (with completely different coefficients for the ``D" and ``H" EoS models). More specific, for the ``D" hybrid model, the change in slope in the curves exemplifies the change in degrees of freedom going from a pure hadronic phase to a phase containing a mixture of hadrons and quarks with increasing quark content. For the ``H" quark model, the number density drops more sharply close to the star surface.

In any case, we do not provide a numerical fit for the magnetic field distribution in the polar direction as a function of baryon number density, as it is evident that this would be model dependent (as far as different degrees of freedom are taken into account) and the fit would have completely different coefficients for each model EoS. Note that in the case of a first order phase transition without the inclusion of a mixed phase (not shown in this work), the difference among models would be even more extreme. As already discussed in Ref.~\cite{Dexheimer:2016yqu}, we do not provide profiles for the magnetic field strength in the stellar equatorial direction, as those are more complicated and current dependent.

\section{Conclusions}

In this work, we provide a numerical fit that allows one to include a magnetic field profile respective to the stellar polar direction in any equation of state in a simple way. This will allow analyses of magnetic field effects in specific models studying, for example,  changes in stiffness, changes in population, phase transitions, temperature, transport properties, etc. A further inclusion of the obtained equations of state in a symmetric static isotropic solution for Einstein's equations (TOV \cite{Tolman:1939jz,Oppenheimer:1939ne}) to obtain macroscopic star properties is not a realistic approach when dealing with strong magnetic fields (\cite{rosana}). This is because the magnetic field distribution is different and more complicated in other directions of the star and the pure magnetic field contribution would have to be added in an isotropic manner, being either positive or negative. In reality, this contribution has different signs in different directions and, therefore, requires a more advanced formalism (such as the one used in this work) which solves Einstein-Maxwell's field equations self-consistently.

This numerical fit for the magnetic field strength was given as a function of baryon chemical potential and is, to a large extent, model independent. This would not be the case for a numerical fit as a function of the baryon number density, in which case the shape of the resulting curves depends substantially on the model. In this work, we have used three very different state-of-the-art equation of state models, built with different assumptions and including different degrees of freedom. They were combined with the solutions of the Einstein-Maxwell's equations in a self-consistent way, in order to provide a formula to calculate how the magnetic field varies with baryon chemical potential, depending only on the dipole magnetic moment of choice and the stellar baryonic mass. The resulting fit is quadratic in form and not exponential as previously assumed. This result is particularly important because it shows that a star with a surface magnetic field of $10^{15}$ G cannot reach a central one of $10^{18}$ G, as previously assumed. Our fit is presented for the two most relevant types of neutron stars, with gravitational masses around $2$ and $1.4$ M$_\odot$.

The authors acknowledge support from NewCompStar COST Action MP1304 and from the LOEWE program HIC for FAIR. Work partially financed by CNPq under grants 308828/2013-5 (R.L.S.F) and 307458/2013-0 (S.S.A).

\section*{References}

\bibliography{wileyarticle-template}%

\end{document}